\documentclass[11pt]{article}
\usepackage{moriond,epsfig}

\def\Journal#1#2#3#4{{#1} {\bf #2}, #3 (#4)}

\def\NIM{\em Nucl. Instrum. Methods}

\def\NPB{{\em Nucl. Phys.} B}

\def\PRL{\em Phys. Rev. Lett.}
\def\PRD{{\em Phys. Rev.} D}

\def\JPG{{\em Jour. Phys.} G}

\def\be{\begin{equation}}
\def\ee{\end{equation}}
\def\bea{\begin{eqnarray}}
\def\eea{\end{eqnarray}}

\begin{document}
\vspace*{4cm}
\title{FACTORISATION FITS AND THE UNITARITY TRIANGLE}

\author{ N. de Groot}
\address{Rutherford Appleton Laboratory, \\Chilton, Didcot OX11 0QX, U.K.}
\author{ W. N. Cottingham}
\address{H.H. Wills Physics Laboratory, Bristol University,\\
Tyndall Avenue,\\ Bristol BS8 1TL, U.K.}
\author{ I. B Whittingham}
\address{School of Mathematical and Physical Sciences,\\ 
            James Cook University,\\ Townsville, Australia 4811}

\maketitle\abstracts{
In this paper we present fits to charmless hadronic B decay data from 
the BaBar, Belle and Cleo experiments using models by Beneke {\em et. al.}
~\cite{BBNS,BBNS2} and Ciuchini {\em et al.}~\cite{CPg}. 
When we include the data
from pseudo-scalar vector decays (PV) the current experimental results favour
the inclusions of a so-called "charming penguin" term. We also present fit
results for the Unitary Triangle parameters and the CP violating asymmetries
}

\section{Introduction}
A wealth of experimental data on hadronic charmless $B$ decays has become
available from the BaBar and Belle experiments. In this paper we present an
analysis, based upon QCD factorisation, of data on hadronic charmless $B$
decays. We also investigate the potential contribution to the decay amplitudes 
of $b$ quark annihilation and of so-called charming penguins. The data we 
attempt to fit includes pseudo-scalar ($\pi$ and $K$) and vector ($\rho$, 
$\omega$, $K^{*}$ and $\phi$) mesons. This is an extension of an 
earlier study~\cite{CWGW}.

The decay amplitude can be described by:
\begin{equation}
A(B \rightarrow M_1 M_2) = \frac{G_f}{\sqrt 2 } 
    \sum_{p=u,c} \sum_{i=1..10} \lambda_p a_i^p <M_1 M_2|Q_i|B>_{fact}
\end{equation}
with $\lambda_p$ the appropriate CKM matrix factor for the process. The
coefficients $a_i$ can be written as: $a_i = a_{i,I} + a_{i,II}$.

The $a_{i,I} $ coefficients are similar to the ones of naive factorisation
with vertex and penguin corrections included. We assume that all the 
light mesons have the same spatial wave function. As a 
consequence the $a_{i,I}$ coefficients are universal. The $a_{i,II}$
coefficients contain the hard spectator interactions. They contain an
end-point divergence from the low energy contributions to the loop integrals,
which is parametrised by the complex parameter $X_H$. The magnitude of $X_H$
is expected to be between 0 and 3. 

$B$ meson decay can also be initiated by the $b$ quark annihilating with
its partner. Contributions to the decay amplitude should be small, of
order $\Lambda_{QCD}/M_B$, compared to the leading $b$ quark decay.
Because of the heavy $b$ quark mass it is expected that perturbative QCD
calculations will give a reliable estimate of the contributions of 
annihilation to the decay amplitude. Apart from the electroweak meson
decay constants, $f_B$, $f_{\pi}$, $f_{\rho}$ etc, there is a low energy
contribution coming from the loop integrals. This is parametrised by the
complex parameter $X_A$. Formula for the annihilation contributions are
given by BBNS~\cite{BBNS2}, we use the extension to pseudo-scalar vector
decays by Du {\em et al.}~\cite{Du}.

Charming penguins orginate from diagrams with a charm quark loop. According
to Ciuchini {\em et al.} they can be enhanced~\cite{CPg}. 
They behave similar to 
the $a_4$ coefficients in signs and Clebsch-Gordan coefficients and take the
following form:
\begin{equation}
A_{b \rightarrow q} = \frac{G_f}{\sqrt 2 }
     [ - V_{uq}^* V_{ub} (\bar P_1 - \bar P_1^{\mbox{\tiny GIM}}) 
       - V_{cq}^* V_{qb} {\bar P_1} ]
\end{equation}

\section{Fitting Method}
\label{sec:fit}
We have attempted to fit the theoretical expressions for branching ratios
with the available data as averaged by the Heavy Flavour Averaging 
Group~\cite{HFAG}. Measured branching ratios for seventeen 
channels are shown in Table~\ref{tab:br}.
We take the measured branching ratios to be the mean of the $B$ and 
$\bar{B}$ decays. The CP asymmetries are not included in the fit. 
Their measurements are not always consistent between the experiments and 
the errors are large. Therefore we prefer to compare the measured results 
with the prediction from the fit.

For convenience we assign to each channel $(h_{1}h_{2})$ a number $\alpha $.
The statistical and systematic errors have been combined into a single error
$\sigma_{\alpha}$. The systematic errors are small in general and we ignore
any correlation between the channels. We then write for a 
$\chi ^{2}$ function
\begin{equation}
\label{fact52}
\chi ^{2}(P_{i}) = \sum_{\alpha} \left[ \left({\rm Br}_{\alpha} (P_{i})-
{\rm Br}_{\alpha} ({\rm exp})\right)/\sigma_{\alpha}\right] ^{2}
+ {\rm additional\;constraints}.
\end{equation}
${\rm Br}_{\alpha}(P_{i})$ are the theoretical branching ratios
in terms of ten parameters $P_{i}, i=1,\ldots ,10 $ which
we take to be the three Wolfenstein CKM parameters $\{A, \rho, \eta\}$ and the
seven soft QCD parameters $ \{ R^{\pi}_{\chi}, R^{K}_{\chi},F_{\pi}, F_{K},
A_{\rho},A_{\omega},A_{K^{*}}\}$. For the fit to the charming penguin model
we introduce four more terms: $\bar P^1$ and $\bar P^{\mbox{\tiny GIM}}$ 
and their phases. In this case we fix $A$ to the world average on 0.82 
and keep $R^{\pi,K}_{\chi}$ fixed
at 1.0.  The well known decay parameters $\{f_{\pi},f_{K},f_{\rho},
f_{\omega},f_{\phi},f_{K^{*}}\}$ are held at their
mean values and the Wolfenstein CKM $\lambda $ parameter is taken to be
$\lambda = 0.2205$. For the divergence parameters we take $|X_H| = |X_A| = 2$,
which leads to reasonable values of e.g. $\sin 2 \beta$.
Additional terms were included in the $\chi ^{2}$
to take into account experimental and theoretical constraints from outside
the data on $B$ decay branching ratios.
We search for a minimum of $\chi ^{2}$ as a function of the $P_{i}$.
We use the MINUIT~\cite{MINUIT} program to minimise the 
$\chi ^{2}$.
The theoretical branching ratios and contributions of the individual channels
to $\chi^{2}$ based on these best fit values are given in Table~\ref{tab:br}.

\begin{figure}[htb]
\begin{center}
\epsfig{figure=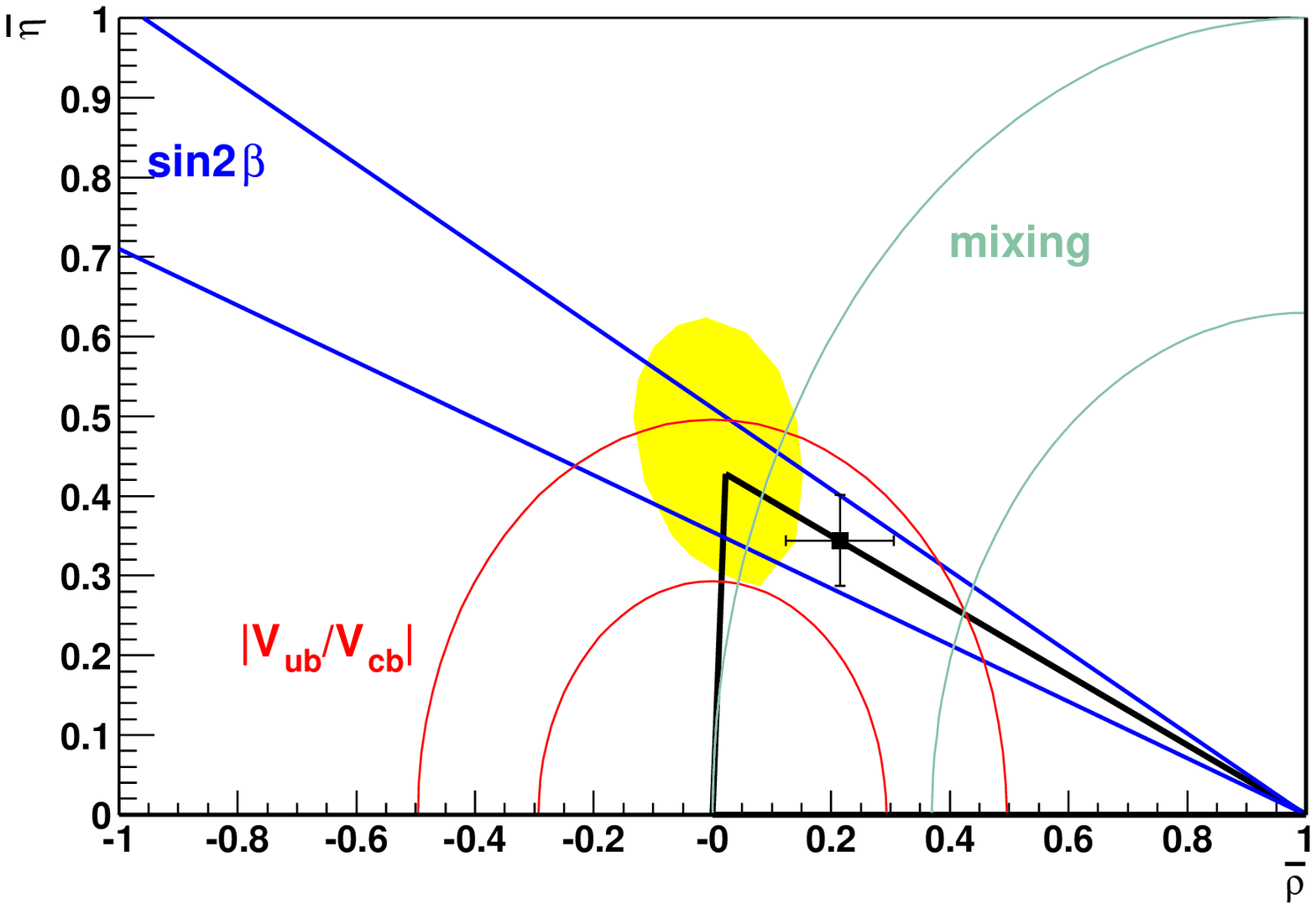,height=1.9in}
\epsfig{figure=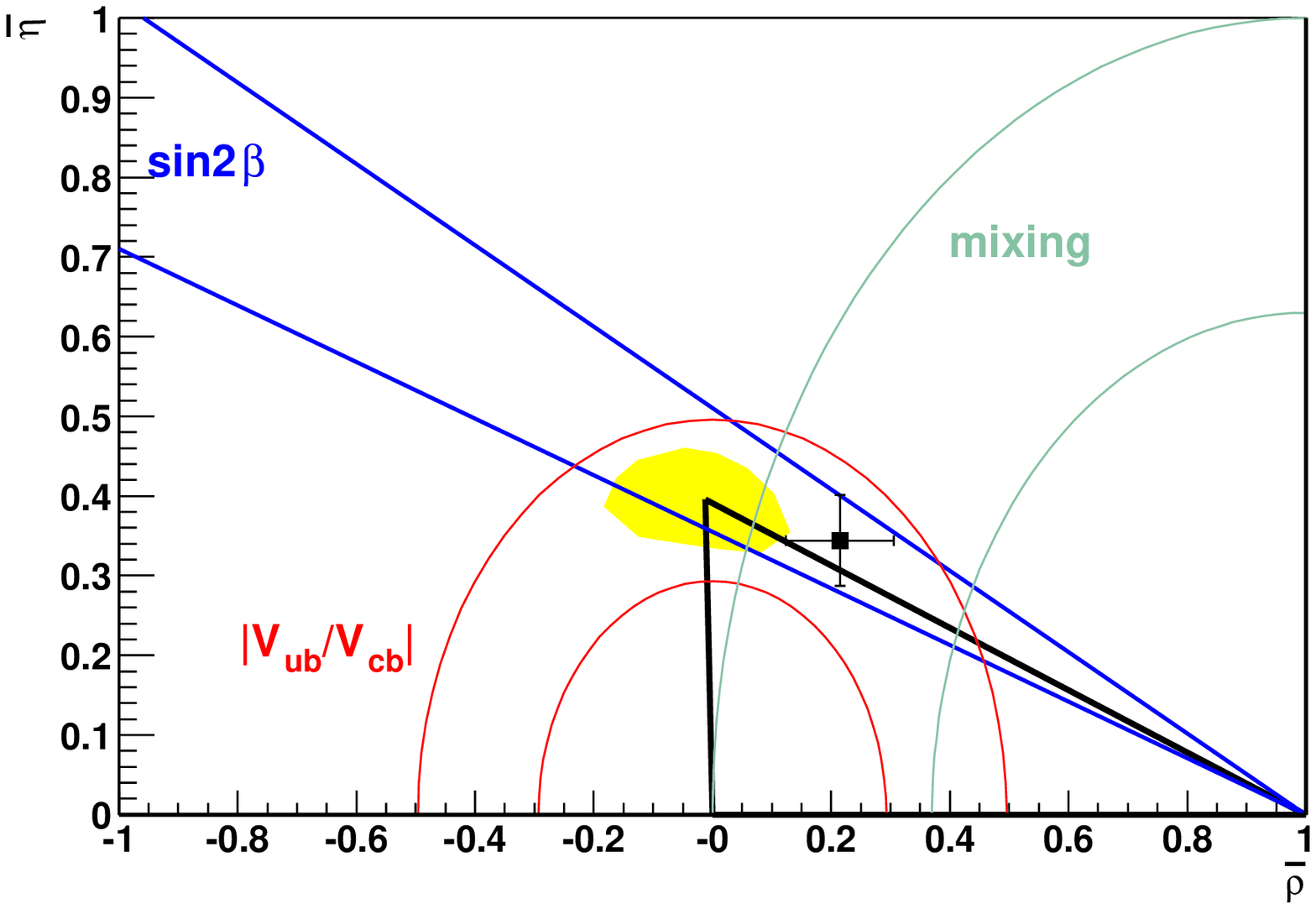,height=1.9in}
\caption{\label{fig:UT}
Result for the Unitarity Triangle fit for BBNS (left) and
charming penguins (right). The shaded area shows the $3\sigma$ allowed region
for the apex of the Unitarity triangle. The data point shows the fit result
from the Unitarity Triangle fit from other measurements taken 
from \em{CKMfitter}$^9$.
}
\end{center}
\end{figure}

The values of the best fit parameters are shown in Table~\ref{tab:fit} 
together with our estimates of the errors.  These errors are of 
course highly correlated. A plot of the error matrix ellipse for the 
Wolfenstein parameters $\rho$ and $\eta$ is shown in Fig.~\ref{fig:UT}. 
For both models, the results in Table~\ref{tab:fit} for the best fit 
values of the various form factors lie within the spread of theoretical 
estimates. The $\chi^2/{\mbox{dof}}$ is 34/16 for the BBNS fit and 16/12
for the charming penguins.

\begin{table}[htb]
\caption{Measured branching ratio Br(exp), experimental error $\sigma $
(errors added in quadrature), theoretical branching ratio for best
fit parameters and contribution to $\chi^{2}$ for various $B$
decay channels. Units are $10^{-6}$.}
\label{tab:br}
\begin{center}
\begin{tabular}{@{}|l|ll|ll|ll|}
\hline
Decay & Br(exp) & $ \sigma  $ 
& BBNS Br & $\chi^{2}$  
& CP Br & $\chi^{2}$  \\
\hline
$\pi^{+}\pi^{-}$ & 4.8  & 0.5  & 5.1 & 0.3  & 4.9  & 0.1  \\
$\pi^{0}\pi^{-}$ & 5.9  & 1.0  & 4.1 & 3.1  & 5.9  & 0.0  \\
$\rho^{\pm}\pi^{\mp}$  & 25.4  & 4.2  & 24.6 & 0.1 & 22.8  & 0.4  \\
$\rho^{0}\pi^{-}$  & 9.6  & 2.0  & 9.3 & 0.1 & 10.0  & 0.1  \\
$\omega \pi^{-}$  & 6.4  & 1.3  & 7.5 & 0.7 & 6.3  &  0.0  \\
$\pi^{0}K^{-}$  & 12.7  & 1.2  & 13.1 & 0.1 & 12.9  &  0.0  \\
$\pi^{-}K^{0}$  & 18.0  & 1.7  & 19.9 & 1.3 & 19.6  & 0.9  \\
$\pi^{-}K^{*0}$  & 12.3  & 2.6  & 4.3 & 9.3  & 8.0  &  2.8  \\
$\omega K^{-}$  & 3.1  & 1.0  & 2.3 & 0.6 & 4.2  & 1.2  \\
$\phi K^{-} $  & 8.9  & 1.0  &  8.6 & 0.1 & 8.7  &  0.1  \\
$\phi K^{*-}  $  & 10.7  & 2.6 &  10.2 & 0.0 & 11.4  & 0.1  \\
$\pi^{-}K^{+}$  & 18.5  & 1.0  & 18.8 & 0.1 &  18.7  & 0.1  \\
$\pi^{-}K^{*+} $  & 12  & 6  &  4.6 & 3.6 & 8.2  & 1.7  \\
$\pi^{0}K^{0} $  & 10.3  & 1.5  & 6.9 & 5.0 & 6.9  & 5.0  \\
$\omega K^{0}$  & 5.9  & 1.9  & 1.5 & 5.2 & 3.5  &  1.6  \\
$\phi K^{0}$  & 10.7  & 2.7  &  8.2 & 0.2 & 11.4  & 0.1  \\
$\phi K^{*0}$  & 8.8  & 1.3  & 9.6 & 0.3 & 8.3  & 0.2  \\
\hline
\end{tabular}\\
\end{center}
\end{table}

\begin{table}[htb]
\caption{Best fit values and one-standard deviation errors for both
methods. $F_{\pi,K}$ and $A_{\rho,\omega,K^*}$ are the transition form
factors, $F^{B \rightarrow M}(0)$. $R^{\pi,K}_{\chi}$ are the chiral 
enhancement factors, which are nominally power supressed, but in practice
$O(1)$.
\label{tab:fit}}
\vspace{0.4cm}
\begin{center}
\begin{tabular}{@{}|l|lllll|}
\hline
& $ F_{\pi}$ & $F_{K}$ & $A_{\rho} $ & $A_{\omega } $ & $ A_{K^{*}}$ \\
\hline
BBNS & $0.243 \pm 0.038 $ & $0.368\pm 0.031$ & $0.326\pm 0.088 $ & 
$ 0.301 \pm 0.086$ & $ 0.314 \pm 0.129 $  \\
CP & $0.285 \pm 0.017 $ & $0.382\pm 0.037$ & $0.328\pm 0.052 $ & 
$ 0.316 \pm 0.064$ & $ 0.322 \pm 0.111 $  \\
& &&&& \\
\hline
& $R^{\pi}_{\chi} $ & $ R^{K}_{\chi}$ & $A$ & $\bar \rho $ & $\bar \eta $  \\
\hline
BBNS & $1.09 \pm 0.24 $ & $ 1.19 \pm 0.19 $ & $ 0.885 \pm 0.061$ &
$ 0.026 \pm 0.069 $  &  $0.366 \pm 0.073 $  \\
CP & {\em 1.0}  &  {\em 1.0}  &  {\em 0.82}  &
$ 0.006 \pm 0.087 $  &  $0.411 \pm 0.039 $  \\
& &&&& \\
\hline
& $\bar P^{1} $ & $ arg(\bar P^{1})$ & 
$\bar P^{\mbox{\tiny GIM}}$ & $arg(\bar P^{\mbox{\tiny GIM}}) $ & \\
\hline
CP & $0.059 \pm 0.007$ & $1.58 \pm 0.10$ & $0.33 \pm 0.12$ & $0.93 \pm 0.21$ &\\ 
\hline
\end{tabular}
\end{center}
\end{table}

\begin{table}[htb]
\caption{CP asymmetries from BaBar and Belle and fit predictions.
Only statistical errors are included.
\label{tab:asys}}
\vspace{0.4cm}
\begin{center}
\begin{tabular}{@{}|l|llll|}
\hline
& BaBar & Belle & BBNS & CP \\
\hline
$A_{CP}$ & & & & \\
$K^+ \pi^- $ & $-0.102 \pm 0.05$ & $-0.07 \pm 0.06$ & $ 0.08 $ & $ 0.00$ \\
$K^+ \pi^0 $ & $-0.09 \pm 0.09$ & $0.23 \pm 0.11$ & $ 0.14 $ & $ 0.06$ \\
$K^0 \pi^+ $ & $-0.17 \pm 0.10$ & $0.07 \pm 0.09$ & $ 0.01 $ & $ 0.11$ \\
$\pi^+ \pi^0 $ & $-0.03 \pm 0.18$ & $-0.14 \pm 0.24$ & $ 0.0 $ & $ 0.0$ \\
$\rho^+ \pi^- $ & $-0.22 \pm 0.08$ & -  & $ -0.04 $ & $ 0.25$ \\
$\rho^+ K^- $ & $0.28 \pm 0.17$ & -  & $ -0.16 $ & $ -0.47$ \\
\hline
$C_{\pi\pi}$ & $-0.30 \pm 0.25$ & $-0.77 \pm 0.27 $& $0.17$ & $-0.03$ \\
$S_{\pi\pi}$ & $+0.02 \pm 0.34$ & $ -1.23 \pm 0.41 $ & $0.10$ & $0.03$ \\
\hline
$C_{\rho\pi}$ & $+0.36 \pm 0.15$ & - &  $0.02$ & $-0.27$ \\
$S_{\rho\pi}$ & $+0.19 \pm 0.24$ & - &  $0.32$ & $0.27$ \\
$\Delta C_{\rho\pi}$ & $+0.28 \pm 0.19$ & - &  $0.17$ & $0.33$ \\
$\Delta S_{\rho\pi}$ & $+0.15 \pm 0.25$ & - &  $0.0$ & $0.05$ \\
\hline
$C_{\phi K^0_S}$ & $-0.80 \pm 0.38$ & $0.56 \pm 0.41 $& $-0.02$ & $-0.21$ \\
$S_{\phi K^0_S}$ & $-0.18 \pm 0.51$ & $ -0.73 \pm 0.64 $ & $0.76$ & $0.74$ \\
\hline
\end{tabular}
\end{center}
\end{table}

\section{Conclusions}
The first conclusion would be that the factorisation approach works well.
Most of the branching fractions in Table~\ref{tab:br} are predicted correctly
by both models. The fitted parameters in Table~\ref{tab:fit} look reasonable 
for both fits. The $\chi^2$ for the charming penguin fit is significantly 
better. This is due to the poor fit of BBNS for the decay modes with a
$K^*$ meson. Also the experimental value for $\omega K^0$ is not easily
accommodated within the BBNS model. Both models have a problem fitting the
$\pi^0 K^0$ mode. Figure~\ref{fig:UT} shows the position of the apex of the
Unitarity Triangle for both fits. The results agree with each other on the
angle $\gamma$, but give a larger value than the fits from other measurements.
The angle $\beta$ agrees well, although it has to be pointed out that for
the BBNS the value of $\beta$ is sensitive to the choice for $X_A$.

Regarding the asymmetries it is too early to come to a conclusion. In many 
cases the experiments do not agree, in others the errors are so large that
a meaningful discrimination is not possible. The asymmetries are in principle
very sensitive to the different models and with improved statistics could
become the final test of factorisation.

\end{document}